\documentclass[reprint,
superscriptaddress,
 amsmath,
 amssymb,
 aps,
 prl
]{revtex4-2}

\usepackage{graphicx}
\usepackage{dcolumn}
\usepackage{bm}
\usepackage{hyperref}
\usepackage{xcolor} 
\usepackage{caption}
\usepackage{subcaption}
\usepackage{verbatim}
 
\begin{document}

\preprint{APS/123-QED}

\title{Continuous variable entanglement in a cold-atoms \\ mirrorless optical parametric oscillator}

\author{G. C. Borba}
\affiliation{%
 Departamento de F{\'i}sica, Universidade Federal de Pernambuco, 50670-901 Recife, PE, Brazil
}
\affiliation{ Instituto de F{\'i}sica da Universidade de São Paulo, P.O. Box 66318, 05315-970 São Paulo, Brazil
}
\author{R. S. N. Moreira}%
\affiliation{%
 Departamento de F{\'i}sica, Universidade Federal de Pernambuco, 50670-901 Recife, PE, Brazil
}%
\affiliation{%
Centro de Ciências, Tecnologia e Saúde, Universidade Estadual da Paraíba, 58233-000 Araruna, PB, Brazil
}

\author{L. S. Cruz}
\affiliation{Centro de Ciências Naturais e Humanas, Universidade Federal do ABC, Santo André, São Paulo, 09210-170, Brazil
}%

\author{M. Martinelli}
\affiliation{ Instituto de F{\'i}sica da Universidade de São Paulo, P.O. Box 66318, 05315-970 São Paulo, Brazil
}%

\author{D. Felinto}
\affiliation{%
 Departamento de F{\'i}sica, Universidade Federal de Pernambuco, 50670-901 Recife, PE, Brazil
}

\author{J. W. R. Tabosa}
\affiliation{%
 Departamento de F{\'i}sica , Universidade Federal de Pernambuco, 50670-901 Recife, PE, Brazil
}
\email{jose.tabosa@ufpe.br}

\date{\today}

\begin{abstract}
In this work, we explore both the internal and external atomic degrees of freedom to observe quantum entanglement between the modes produced by a mirrorless optical parametric oscillator operating below the oscillation threshold in a sample of free-space cold cesium atoms. 
Using a new heterodyne technique, we recover the covariance matrix that reveals the quantum entanglement for two different pairs of modes, thus demonstrating the generation of four entangled modes in this system. Applications to quantum networks, and the possibilities of studying higher orders of entanglement are a direct consequence of the present study.

\end{abstract}

\maketitle

A nonlinear optical medium can efficiently generate forward and backward waves through parametric processes, conserving energy and momentum between input and output fields. The first proposal to use this principle to achieve mirrorless optical oscillation, almost sixty years ago, considered a three-wave-mixing process in a medium with second-order $\chi^{(2)}$ nonlinearity~\cite{Harris1966}. The implementation of this original idea succeeded only in 2007~\cite{Canalias2007}, using quasi phase-matching in a nonlinear photonic structure to satisfy its stringent phase-matching requirements. In 1977, a decade after the original proposal, it was shown that the four-wave mixing (FWM) process, employing a third-order $\chi^{(3)}$ nonlinearity, could also result in mirrorless optical oscillation~\cite{Yariv1977}. Since the phase-matching requirements were much easier to achieve in this case, in the following year, the first experimental observation of mirrorless oscillation in four-wave mixing was already reported~\cite{Pepper1978}. Since then, in the last four decades, many groups have experimentally observed the formation of a mirrorless optical parametric oscillator (MOPO) in four-wave mixing in different physical systems~\cite{TanNo1980,Odulov1984,Leite1986,Odoulov1993,Neumann1995,Zibrov1999,Uesu2004,Harada2007,Greenberg2012,Zhang2014,Mei2017,Lopez2019}. The MOPO based on both kinds of nonlinearities may work as robust and self-aligned coherent light sources analogous to distributed feedback lasers~\cite{Canalias2007,Odulov1988,Khurgin2007}. In addition, the $\chi^{(2)}$ MOPO provides tunable coherent light sources covering wavelengths in the terahertz~\cite{Nawata2019} and near-infrared regions~\cite{Molster2022}, while the less-stringent phase-matching condition of the $\chi^{(3)}$ MOPO is well suited for the self-established formation of cascaded parametric oscillations in multiple directions~\cite{TanNo1980,Grynberg1988a, Grynberg1988b, Odoulov1993,Neumann1995,Harada2007,Lopez2019}.

On the other hand, optical parametric oscillators (OPOs) have been established as a standard platform to generate non-classical states in continuous variables (CV). From the squeezed vacuum in sub-threshold operation \cite{Kimble1986} to the twin beam generation \cite{Fabre1987}, and the entanglement generation below \cite{Kimble1992} and above \cite{Villar2005} threshold, these systems soon found applications in construction of networks of entangled states, directly coupling distinct modes, discriminated either in time \cite{furusawacluster}, frequency \cite{pfistercluster} or wavelength \cite{trepscluster}. Use of third-order nonlinearity, on the other hand, led to the possibility of integrating those devices, resulting in on-chip manipulation of quantum states that led to demonstrations of quantum advantage \cite{Madsen2022}.

We show here that a mirrorless OPO based on the four-wave mixing process in a cold atomic cloud \cite{Lopez2019} can produce entangled fields from the parametric amplification of vacuum input. Differently from usual FWM experiments involving only two beams \cite{Boyer2008, Marino2009}, the process in MOPO is intrinsically multipartite. Using a new approach that allows us the simultaneous measurement of both quadratures of the field, at the cost of an extra unit of vacuum noise, we demonstrate entanglement between three fields in two independent measurements. However, the symmetry of the system indicates a more extensive network of entangled states that could be useful for quantum information applications. Moreover, this MOPO has been used to store and recover the orbital angular momentum of a light field \cite{Lopez2020} in the atomic degrees of freedom, suggesting a possibility of combining multipartite entanglement in CV with memory capabilities.

\begin{figure}[h]
\begin{center}
  \begin{subfigure}[b]{0.45\textwidth}
    \includegraphics[width=\linewidth]{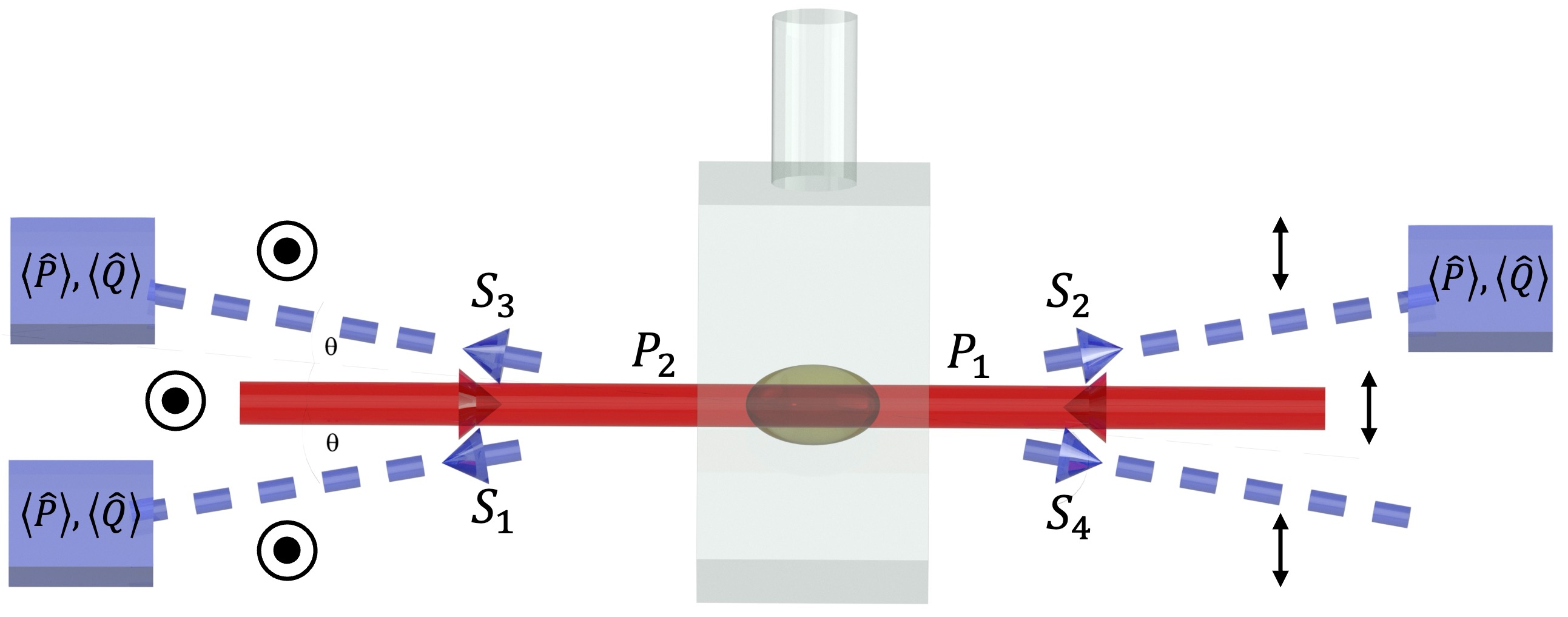}
    \caption{} \label{fig:1a}
  \end{subfigure}%
  \vspace*{\fill}
  \begin{subfigure}[b]{0.4\textwidth}
    \centering
    \includegraphics[width=\linewidth]{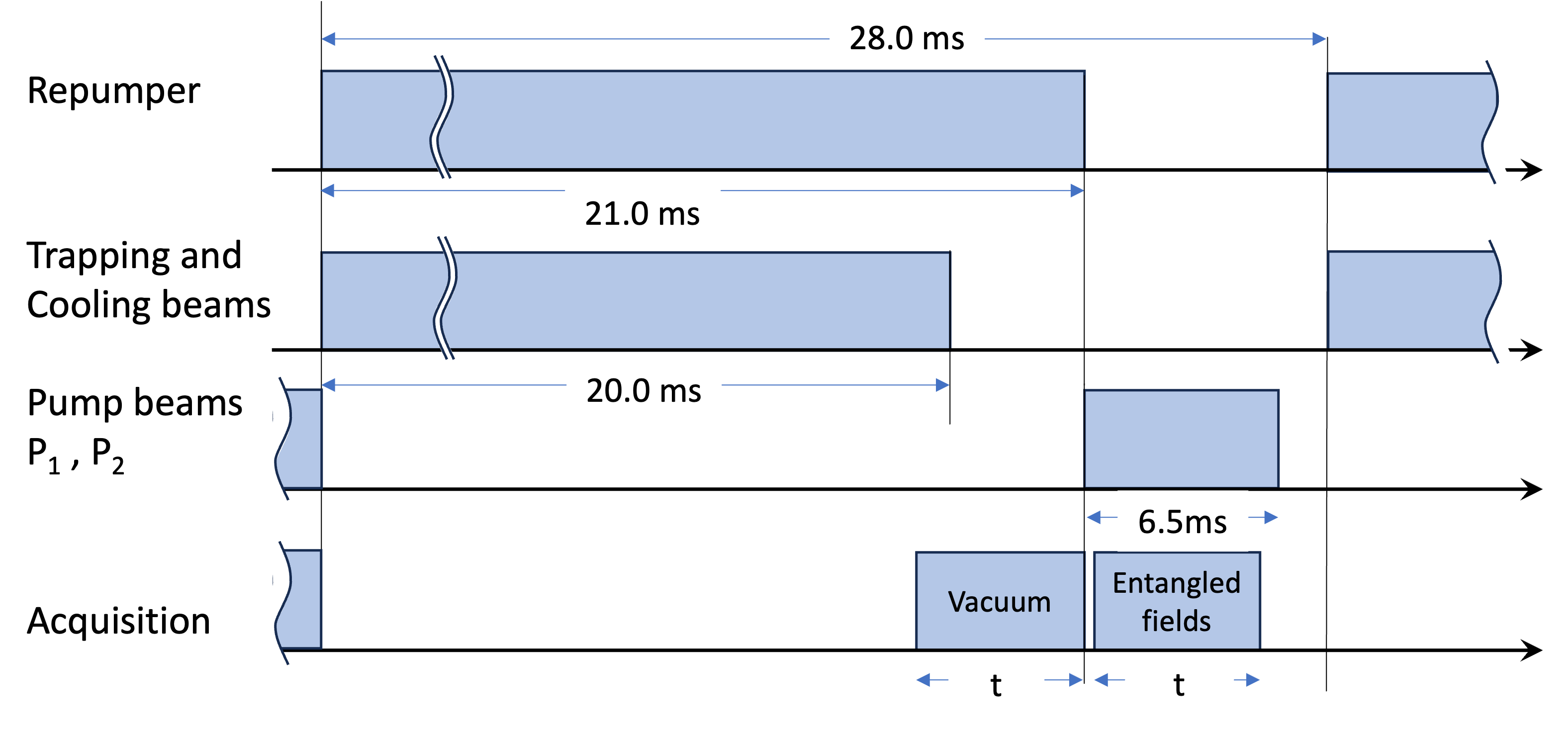}
    \caption{} \label{fig:1b}
  \end{subfigure}
 \end{center}
\caption{(a) Simplified experimental setup. The counter-propagating pumping beams $\left(P_1\right.$ and $\left.P_2\right)$ with linear orthogonal polarizations interact with the atoms and generate the four modes specified by $S_1, S_2, S_3$, and $S_4$. Modes $S_1, S_2$ and $S_3$ are sent to independent heterodyne detection systems.
(b) Time sequence of the involved fields and detection process.} 
\label{fig:spatial_configuration}
\end{figure}

The MOPO involves two counter-propagating pumping beams ($P_1$ and $P_2$) with the same frequency $\omega$ and orthogonal linear polarizations, as depicted in Fig. \ref{fig:spatial_configuration}, over a cloud of cold Cesium atoms.  For such beam geometry interacting nearly in resonance with a degenerate two-level system, multiple FWM processes can take place associated with Zeeman coherence as well as coherences between momentum states via recoil-induced resonance (RIR) \cite{sup_mat}. In these conditions, it has been demonstrated that a probe field can be amplified by a factor exceeding 2000 due a feedback loop \cite{Lopez2019}.

The dynamic of the pairwise photon creation involving the four modes can be understood with a combination of parametric amplifiers, and the Hamiltonian of this process can be written as
\begin{align}
\hat{H}_I= i &\hbar \big[\chi_{12}\hat{a}_1 \hat{a}_2+\chi_{34}\hat{a}_3 \hat{a}_4  \nonumber \\ 
  &+\chi_{24}\hat{a}_2 \hat{a}_4+\chi_{13}\hat{a}_1 \hat{a}_3 - \text {H.c.}\big],
\label{eq:hamiltonian} 
\end{align}
using the field operators for modes 1-4. A useful approach considering balanced couplings in the FWM process, such that $\chi_{ij}=\chi$, allows for a simple substitution in the operators, as 
$\hat{a}_a=\left(\hat{a}_1+\hat{a}_4\right) / \sqrt{2}$, $\hat{a}_b=\left(\hat{a}_2+\hat{a}_3\right) / \sqrt{2}$, resulting in 
\begin{equation}\label{eq:hamiltonian2}
  \hat{H}_I=i \hbar \chi\left[\hat{a}_a \hat{a}_b - \text { H.c. }\right].
\end{equation}
In this case, one can observe that we have the generation of photon pairs on a mode given by a combination of modes 1 and 4 and another in a combination of modes 2 and 3, corresponding to a Bloch-Messiah decomposition involving two-mode squeezing states \cite{sup_mat}. 
As  first approach to the system, we started by the evaluation of entanglement involving  just two pairs of modes, namely, $S_1-S_2$ and $S_1-S_3$, as shown in Fig.\ref{fig:spatial_configuration}, in two independent measurements.

The experiment is performed in a cold cloud of cesium atoms obtained from a magneto-optical trap (MOT) using the Zeeman structure of the hyperfine levels $6 S_{1 / 2}(F=4)$ and $6 P_{3 / 2}\left(F^{\prime}=5\right)$.
The trap (cooling beam, repump beam, and magnetic quadrupole field) are kept on for 20 ms, then followed by a 1 ms cycle of repump (Fig. \ref{fig:spatial_configuration}), resulting in a cold sample of Cs atoms in the $6 S_{1 / 2}(F=4)$ ground state, with about $2 \mathrm{~mm}$ diameter, optical density of 5, and temperature in hundreds of $\mu K$ range. 
 Three pairs of Helmholtz coils are employed to compensate for residual magnetic fields, whose current values are adjusted by optimization of microwave spectroscopy \cite{Allan2016} in the cesium clock transition $6 S_{1 / 2}(F=3) \rightarrow 6 S_{1 / 2}(F=4)$.

Differently from the configuration in \cite{Lopez2019}, we use a probe beam only for alignment purposes \cite{sup_mat} and only measure the spontaneous process. The pump beams $P_1$ and $P_2$ have approximately the same waist diameter of 1.6 mm and are switched on during 6.5 ms after turning off all MOT beams and trapping magnetic field. Pump beams are provided by a Ti:sapphire laser locked to a saturated absorption signal and their frequencies are shifted by an acoustic-optic modulator detuned around $\Delta = 25 MHz (\approx 5 \Gamma)$ below the resonance frequency of the cesium closed transition $6 S_{1 / 2}, F=4 \rightarrow 6 P_{3 / 2}, F^{\prime}=5$ 

The measurement of the field uses a heterodyne detection based on the interference of the amplified vacuum with an optical field, shifted by 5 MHz from the pump frequency, used as an optical local oscillator in a balanced detection. Subtraction of the photodiode's currents is further demodulated by a pair of in-quadrature electronic references at the shift frequency, thus providing a combined measurement of both quadratures of the signal of interest, plus one unit of vacuum \cite{sup_mat}. The resulting measurement of the quadrature for each field is recorded by an analog-to-digital converter and the statistics is further performed, in order to recover the covariance matrix of the signal.

Differently from the majority of the CV systems, which have a continuous generation of the fields, a measurement involving the cold cloud is performed when the trapping fields are turned off and the cloud is left for free expansion. Given the narrow bandwidth of the RIR resonances (on the order of 20 kHz \cite{Lopez2019}), there is a lower limit for the repetition rate of the aquisitions.
In our case, an acquisition time shorter than 25 $\mu$s implies in correlation of sequential measurements and data redundancy.
On the other hand, the upper limit for the acquisition cycle is given by free-expansion of the cloud. A reasonable time between 3.0 and 6.5 ms results in a total of 120 to 260 acquired values of the quadratures for each run of the MOT, what is enough for a reliable statistics to evaluate the variances of the quadratures for distinct values of the pump beams. Precise calibration of the standard quantum level, used as our normalization factor in the present results, is assured by running a previous cycle of acquisition before turning on the pump fields (Fig. \ref{fig:spatial_configuration}). The resulting input on the detection has only vacuum fluctuations, independent of the trap and repump fields as was experimentally verified.

The experiment is performed in two distinct configurations of the acquisition system. One performs the detection over opposite outputs 1 and 2, while the other evaluates the quadratures of outputs 1 and 3. While performed independently, these results are obtained under the same experimental conditions of the MOPO, and they are used to check the entanglement of these modes.

The entanglement properties of the generated Gaussian states can be completely characterized by their second-order moments, which can be conveniently organized as a covariance matrix. $\mathbb{V} =\left\langle\hat{\xi} \hat{\xi}^T + \left( \hat{\xi} \hat{\xi}^T\right)^T\right\rangle$/2, involving the quadratures of the fields $\hat{\xi}=\left(\hat{P}_i, \hat{Q}_i, \hat{P}_j, \hat{Q}_j\right)^T $. For a given pair of modes, we can divide the covariance matrix into three $2 \times 2$ submatrices, from which two ($\mathbb{A}_j$) represent the reduced covariance matrices of the individual subsystems and one ($\mathbb{C}_{ij}$) expresses the correlations between the subsystems
\begin{equation}\label{eq:covariance}
\mathbb{V}=\left(\begin{array}{cc}
\mathbb{A}_i & \mathbb{C}_{ij} \\
\mathbb{C}_{ij}^T & \mathbb{A}_j
\end{array}\right) .
\end{equation}

Entanglement can be probed by the physicality of the covariance matrix under partial transposition \cite{Simon2000}.
For a pair of modes, the physicallity condition can be recast in terms of the determinant of the covariance matrix and its submatrices as \cite{Simon2000, FBarbosa2011}
\begin{equation}\label{eq: W}
W = 1 + \operatorname{Det}[\mathbb{V}]
 - 2 \operatorname{Det} [\mathbb{C}_{ij}] 
-\operatorname{Det} [\mathbb{A}_i] 
- \operatorname{Det} [\mathbb{A}_j] \geqslant 0 .
\end{equation}
The partial transposition modifies the sign of $\operatorname{Det} [\mathbb{C}_{ij}]$, resulting in the following sufficient and necessary condition for entanglement in bipartite Gaussian states:
\begin{equation}
W_{\mathrm{PPT}}=1+\operatorname{Det}[\mathbb{V}] 
+ 2 \operatorname{Det} [\mathbb{C}_{ij}] 
- \operatorname{Det} [\mathbb{A}_i] 
- \operatorname{Det} [\mathbb{A}_j] < 0 .
\label{eq: Wppt}
\end{equation}

\begin{figure}
\includegraphics[width=\columnwidth]{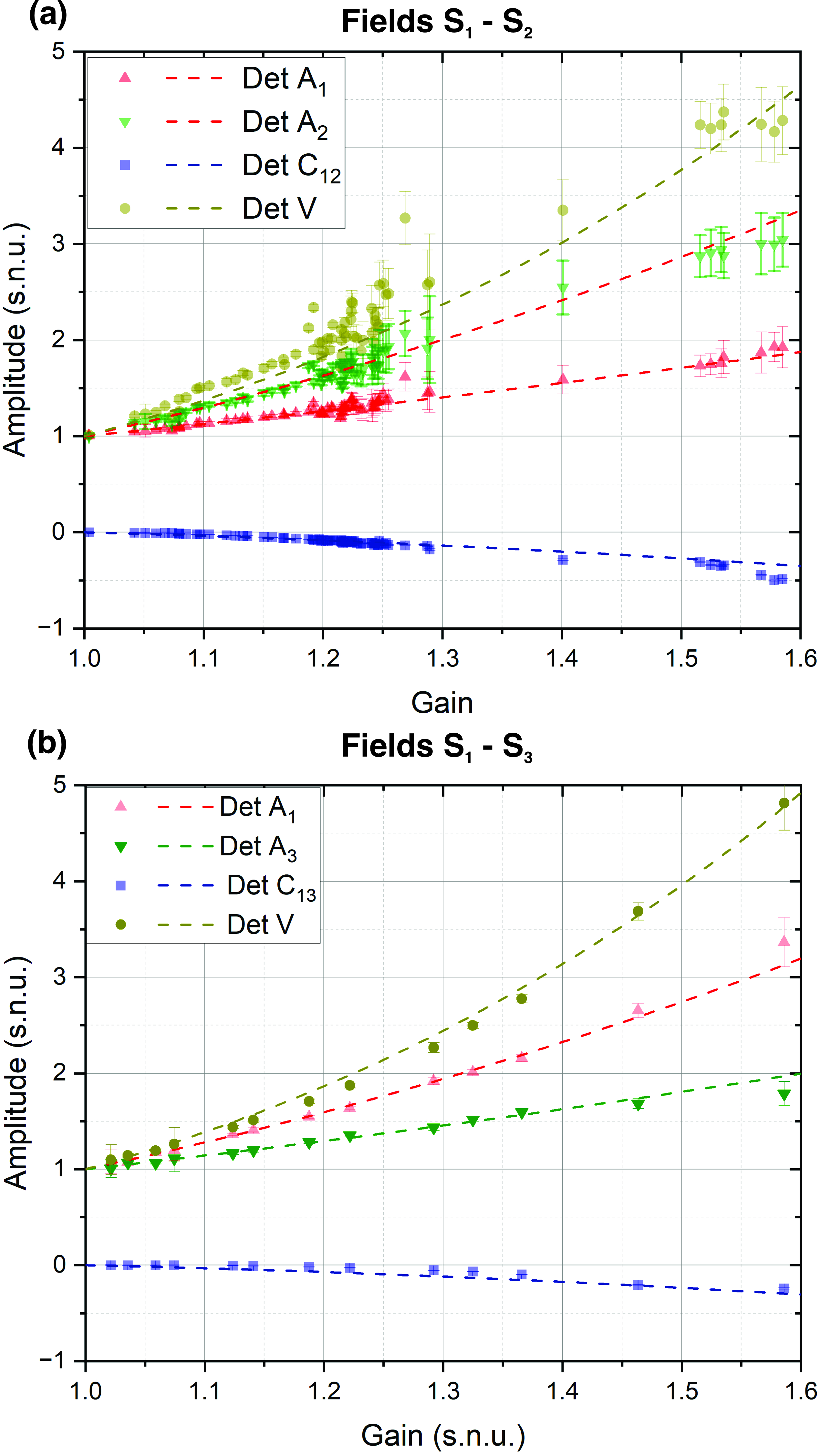}
\caption{\label{fig:determinants} Measured values for the determinants of the covariance matrices of fields $i$ ($\operatorname{Det}[\mathbb{A}_{i}]$) and j ($\operatorname{Det}[\mathbb{A}_j]$), the cross-correlation $\operatorname{Det}[\mathbb{C}_{ij}]$ and the covariance matrix $\operatorname{Det}[V]$, for modes 1 and 2 (a) and modes 1 and 3 (b). The lines are adjustment of the model, as described in the text.}
\end{figure}

The terms of the covariance matrix are evaluated for each run of the experiment, and from them the determinants used in Eqs.~(\ref{eq: W}, \ref{eq: Wppt}) are calculated. Afterwards, these results are grouped according to the measured pump intensity, and the average of the determinants and their standard deviation are calculated. 
The pair of quadratures of each output field is used to evaluate the gain of the MOPO. The four variances are averaged and then normalized to the variance of the vacuum present at the input, obtained from the first part of the acquisition cycle (Fig.\ref{fig:1b}). Figure~\ref{fig:determinants} presents the evaluated determinants over many MOT cycles, for pump intensities up to $1.6$ mW/cm$^2$.

The gain unbalance between the generated modes leads to an unbalance between the determinants of matrices $\mathbb{A}_i$, and is a consequence of the occasional imbalance in the couplings $\chi_{ij}$\cite{sup_mat}.
This behavior can be analyzed, following the description of the process given by Eq. \ref{eq:hamiltonian}, using a Bloch-Messiah decomposition \cite{sup_mat}, and is the only fitting parameter left in the modeling of the curves for these determinants in Fig.\ref{fig:determinants}.
The different relative variance between the fields and the intensity of the correlations are related to their high
sensibility to alignment over the ensemble which changes the phase-matching condition and the competition between the modes. The nature of each FWM process itself can also influence, since the Backward FWM (between $S_1$ and $S_2$) is perfectly phase-matched whereas the Forward FWM (between $S_1$ and $S_3$) is not.

Entanglement comes from the correlation between the modes.
 Notice that $\operatorname{Det} [\mathbb{C}_{ij}] < 0$ over the entire range, fulfilling a necessary condition for entanglement \cite{Simon2000}. We model the expected outcome of this determinant following the Bloch-Messiah decomposition, but with a weighting factor accounting for the mismatch of the spatial modes that are selected by the heterodyne detection.
 The model of the curve for $\operatorname{Det} [\mathbb{C}_{ij}]$ considers a mode mismatch of 0.65 and 0.7 between amplified vacuum modes (1,3) and (1,2), respectively, at the heterodyne measurement, as described in \cite{sup_mat} and is compatible with the observed mismatch coming from a intense generated field upon the injection of a seed through port 1.

\begin{figure}
{\includegraphics[width=\columnwidth]{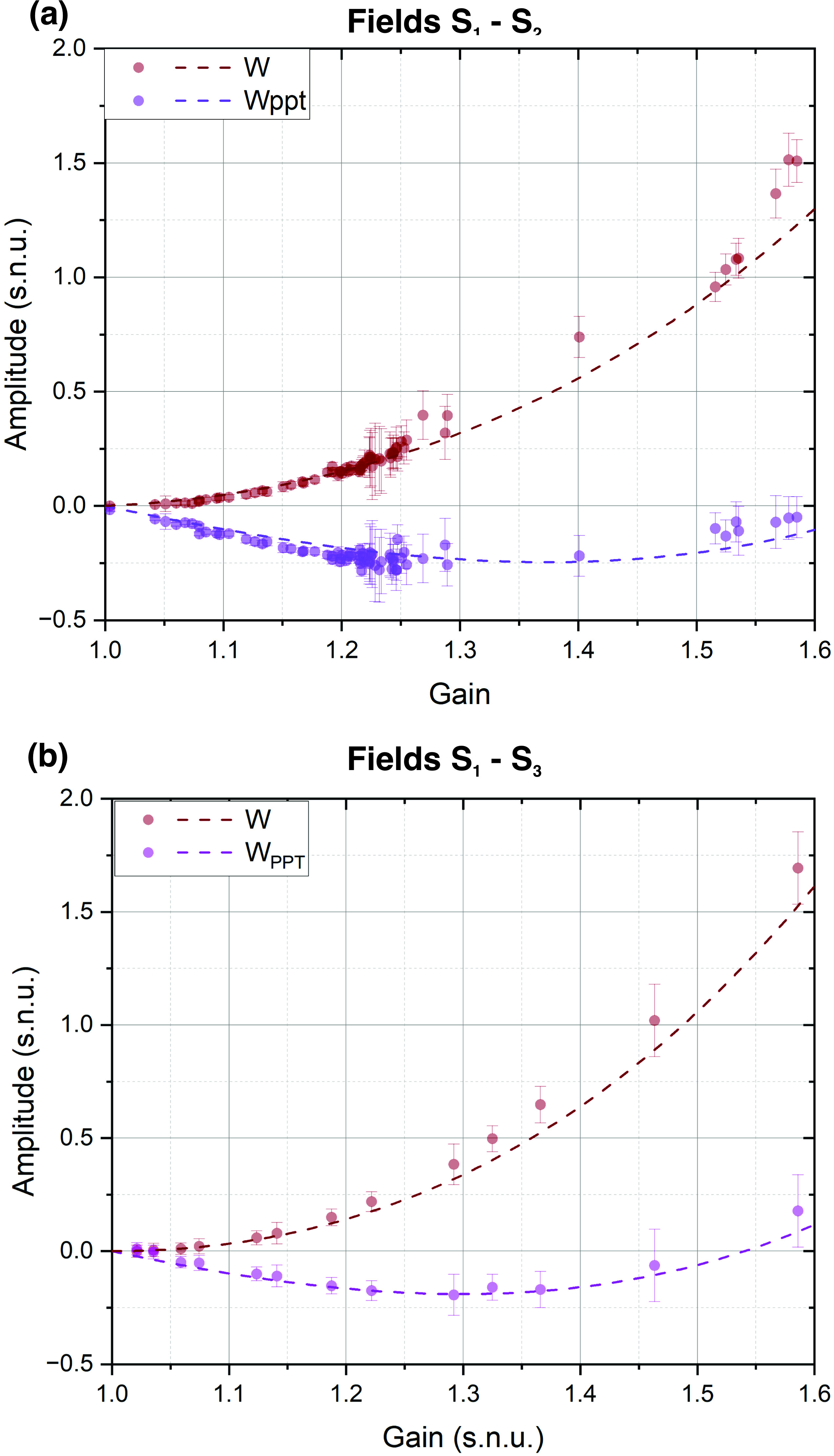}}   
\caption{\label{fig:W_PPT} Physicality test of the measured covariance matrix ($W$) and entanglement witness ($W_{PPT}$), evaluated from the data in Fig. \ref{fig:determinants}, and their respective adjustement (for the lines) } 
\end{figure}

These results allow the evaluation of the entanglement witness $W_{PPT}$, along with the physicality test of the covariance matrix $W$, as presented in Fig. \ref{fig:W_PPT}. Experimental data presents a good match to the calculated results from the model, all of them evaluated from the values in Fig.~\ref{fig:determinants}. We can notice that, for a gain smaller than 1.5, entanglement of the pair of modes is observed. The loss of observed entanglement for higher gains comes from the imperfect matching of the modes on the detection.

The complete symmetry in geometry (Fig. \ref{fig:1a}) and atomic dynamics~\cite{sup_mat} allow us to infer that the correlations between $S_4 - S_3$ and $S_4 - S_2$ should follow a behavior similar to $S_1 - S_2$ and $S_1 - S_3$, respectively. Therefore, even though a full quadripartite characterization requires simultaneous measurement of all four modes, the existence of entanglement in two different sub-partitions and the symmetry of the whole system provides strong evidence of the system’s multipartite nature.

It is our expectation that, under the full measurement of the involved modes, a rich structure of entanglement can thus be recovered, and the observed results, presented here for modes (1,2) and (1,3), will be present in pairwise analyses involving pairs of modes (3,4) and (2,4), as well the several bipartitions of the four modes (1-2,3-4), (1-3,2-4) and so on, as could be measured by technique here developed.

Moreover, as demonstrated in the previous work of the group, the driven atomic cloud has memory \cite{Lopez2020}. Therefore, it should be possible to have this system as a key component in quantum networks, using the possibility to generate large-scale spatially separated entanglement using both spatial and time multiplexing for applications in quantum information.

The MOPO provides an interesting tool for a multi-mode node in a quantum information network. Its reproducible generation in a cloud of cold atoms provides pulses of entangled modes with a seamlessly unbounded parametric gain. Thus, extremely high levels of quantum correlation can be generated in this system. On the other hand, the capacity to couple more than four modes \cite{TanNo1980, Grynberg1988a, Grynberg1988b, Neumann1995,Xiao2009}, can eventually lead to more complex Hamiltonians with the ability to expand this quantum toolbox beyond the bilinear set. The process presented here is only the first step in this investigation.

\section{Acknowledgements}
This study was financed, in part, by the São Paulo Research Foundation (FAPESP), Brasil. Process Number \#2018/20159-8, \#2021/06535-0, \#2022/09436-5; Office of Naval Research (N62909-23-1-2014).
Conselho Nacional de Desenvolvimento Científico e Tecnológico (INCT-IQ 465469/2014-0); Coordenação de Aperfeiçoamento de Pessoal de Nível Superior (PROEX 23038.003069/2022-87); Fundação de Amparo à Ciência e Tecnologia do Estado de Pernambuco (APQ-0616-1.05/18);


\begin{thebibliography}{99}

\bibitem{Harris1966}
S. E. Harris, 
\emph{Proposed backward wave oscillation in the infrared}, \href{http://dx.doi.org/10.1063/1.1754668}{App. Phys. Lett. \textbf{9}, 114-116 (1966)}.

\bibitem{Canalias2007}
C. Canalias and V. Pasiskevicius, 
\emph{Mirrorless optical parametric oscillator}, \href{http://dx.doi.org/10.1038/nphoton.2007.137}{Nat. Photonics \textbf{1}, 459-462 (2007)}.

\bibitem{Yariv1977}
A. Yariv and D. M. Pepper, 
\emph{Amplified reflection, phase conjugation, and oscillation in degenerate four-wave mixing}, \href{https://doi.org/10.1364/OL.1.000016}{Opt. Lett. \textbf{1}, 16-18 (1977)}.

\bibitem{Pepper1978}
D. M. Pepper, D. Fekete, and A. Yariv, 
\emph{Observation of amplified phase-conjugate reflection and optical parametric oscillation by degenerate four-wave mixing in a transparent medium},
\href{https://doi.org/10.1063/1.90185}{Appl. Phys. Lett. \textbf{33}, 41-44 (1978)}.

\bibitem{TanNo1980}
N. Tan-No, T. Hoshimiya, and H. Inaba, 
\emph{Dispersion-Free Amplification and Oscillation in Phase-Conjugate Four-Wave Mixing in an Atomic Vapor Doublet},
\href{https://doi.org/10.1109/JQE.1980.1070455}{IEEE J. Quantum Electron. \textbf{QE-6}, 147-153 (1980)}.

\bibitem{Grynberg1988a}
G. Grynberg,
\emph{Mirrorless four-wave mixing oscillation in atomic vapors},
\href{https://doi.org/10.1016/0030-4018(88)90423-3}{Opt. Comm. \textbf{66},  321-324 (1988)}

\bibitem{Grynberg1988b}
G. Grynberg, E. Le Bihan, P. Verkerk, P. Simoneau, J.R.R. Leite, D. Bloch, S. Le Boiteux, M. Ducloy
Observation of instabilities due to mirrorless four-wave mixing oscillation in sodium,
\href{https://doi.org/10.1016/0030-4018(88)90028-4}{Opt. Comm. \textbf{67},  363-366 (1988)}

\bibitem{Odulov1984}
S. G. Odulov, S. S. Slyusarenko, and M. S. Soskin, 
\emph{Laser with dynamic gratings of free carriers in a semiconductor},
\href{https://doi.org/10.1070/QE1984v014n05ABEH005077}{Sov. J. Quantum Electron. \textbf{14}, 589-590 (1984)}.

\bibitem{Leite1986}
J. R. R. Leite, P. Simoneau, D. Bloch, S. Le Boiteux, and M. Ducloy, 
\emph{Continuous-Wave Phase-Conjugate Self-Oscillation Induced by Na-Vapour Degenerate Four-Wave Mixing With Gain},
\href{https://doi.org/10.1209/0295-5075/2/10/002}{Europhys. Lett. \textbf{2}, 747-753 (1986)}.

\bibitem{Odoulov1993}
S. Odoulov, R. Jungen, and T. Tschudi, 
\emph{Mirrorless Coherent Oscillation Due to Vectorial Four-Wave Mixing in LiNbO$_3$},
\href{https://doi.org/10.1007/BF00325240}{Appl. Phys. B \textbf{56}, 57-61 (1993)}.

\bibitem{Neumann1995}
J. Neumann, G. J\"akel, and E. Kr\"atzig, 
\emph{Holographic scattering lines and mirrorless oscillation in BaTiO$_3$},
\href{https://doi.org/10.1364/OL.20.001530}{Opt. Lett. \textbf{20}, 1530-1532 (1995)}.

\bibitem{Zibrov1999}
A. S. Zibrov, M. D. Lukin, and M. O. Scully, 
\emph{Nondegenerate Parametric Self-Oscillation via Multiwave Mixing in Coherent Atomic Media},
\href{https://doi.org/10.1103/PhysRevLett.83.4049}{Phys. Rev. Lett. \textbf{83}, 4049-4052 (1999)}.

\bibitem{Uesu2004}
Y. Uesu, K. Yasukawa, N. Saito, S. Odoulov, K. Shcherbin, A. I. Ryskin, and A. S. Shcheulin, 
\emph{Backward-wave four-wave mixing and coherent oscillation in CdF$_2$:Ga,Y},
\href{https://doi.org/10.1007/s00340-004-1428-3}{Appl. Phys. B \textbf{78}, 601-605 (2004)}.

\bibitem{Harada2007}
K. Harada, M. Ogata, and M. Mitsunaga, 
\emph{Four-wave parametric oscillation in sodium vapor by electromagnetically induced diffraction},
\href{https://doi.org/10.1364/OL.32.001111}{Opt. Lett. \textbf{32}, 1111-1113 (2007)}.

\bibitem{Greenberg2012}
J. A. Greenberg and D. J. Gauthier, 
\emph{Steady-state, cavityless, multimode superradiance in a cold vapor},
\href{https://doi.org/10.1103/PhysRevA.86.013823}{Phys. Rev. A \textbf{86}, 013823 (2012)}.

\bibitem{Zhang2014}
K. Zhang, J. Guo, C.-H. Yuan, L. Q. Chen, C. Bian, B. Chen, Z. Y. Ou, and W. Zhang, 
\emph{Mirrorless parametric oscillation in an atomic Raman process},
\href{https://doi.org/10.1103/PhysRevA.89.063826}{Phys. Rev. A \textbf{89}, 063826 (2014)}.

\bibitem{Mei2017}
Y. Mei, X. Guo, L. Zhao, and S. Du,
\emph{Mirrorless Optical Parametric Oscillation with Tunable Threshold in Cold Atoms},
\href{https://doi.org/10.1103/PhysRevLett.119.150406}{Phys. Rev. Lett. \textbf{119}, 150406 (2017)}.

\bibitem{Lopez2019}
J. P. Lopez, A. M. G. de Melo, D. Felinto, and J. W. R. Tabosa, 
\emph{Observation of giant gain and coupled parametric oscillations between four optical channels in cascaded four-wave mixing},
\href{https://doi.org/10.1103/PhysRevA.100.023839}{Phys. Rev. A \textbf{100}, 023839 (2019)}.

\bibitem{Odulov1988}
S. G. Odulov, S. S. Slyusarenko, M. S. Soskin, and A. I. Khizhnyak, 
\emph{Mirror-free four-wave parametric oscillation in CdTe crystals},
\href{https://doi.org/10.1070/QE1988v018n08ABEH012386}{Sov. J. Quantum Electron. \textbf{18}, 977-980 (1988)}.

\bibitem{Khurgin2007}
J. B. Khurgin, 
\emph{Mirrorless magic}, 
\href{https://doi.org/10.1038/nphoton.2007.131}{Nature Photon \textbf{1}, 446-447 (2007)}.

\bibitem{Nawata2019}
K. Nawata, Y. Tokizane, Y. Takida, and H. Minamide, 
\emph{Tunable Backward Terahertz-wave Parametric Oscillation}, \href{https://doi.org/10.1038/s41598-018-37068-7}{Sci. Rep. \textbf{9}, 726 (2019)}.

\bibitem{Molster2022}
K. M. Molster, M. Guionie, P. Mutter, J.-B. Dherbecourt, J.-M. Melkonian, X. Delen, A. Zukauskas, C. Canalias, F. Laurell, P. Georges, M. Raybaut, A. Godard, and V. Pasiskevicius, 
\emph{Pump Tunable Mirrorless OPO: an Innovative Concept for Future Space IPDA Emitters},
\href{https://hal.science/hal-03853052}{ICSO (International Conference on Space Optics ) 2022, Oct 2022, Dubrovnik, Croatia.}.

\bibitem{Kimble1986}
L.~A. Wu, H.~J. Kimble, J.~L. Hall, and H.~F. Wu, 
\emph{Generation of squeezed states by parametric down conversion},
\href{https://doi.org/10.1103/PhysRevLett.57.2520}{Phys. Rev. Lett. \textbf{57},~2520--2523, (1986)}.
  
\bibitem{Fabre1987}
A.~Heidmann, R.~J. Horowicz, S.~Reynaud, E.~Giacobino, C.~Fabre, and G.~Camy,
\emph{Observation of quantum noise reduction on twin laser beams},  
\href{https://doi.org/10.1103/PhysRevLett.59.2555}{ Phys. Rev. Lett. \textbf{59}, 255--257 (1987)}.

\bibitem{Kimble1992}
Z.~Y. Ou, S.~F. Pereira, H.~J. Kimble, and K.~C. Peng, 
\emph{Realization of the Einstein-Podolsky-Rosen paradox for continuous variables},
\href{https://doi.org/10.1103/PhysRevLett.68.3663}{Phys. Rev. Lett. \textbf{68}, 3663--3666, (1992)}.

\bibitem{Villar2005}
A. S. Villar, L. S. Cruz, K. N. Cassemiro, M. Martinelli, and P. Nussenzveig, 
\emph{Generation of Bright Two-Color Continuous Variable Entanglement},
\href{https://doi.org/10.1103/PhysRevLett.95.243603}{Phys. Rev. Lett. \textbf{95}, 243603 (2005)}.

\bibitem{furusawacluster}
S. Yokoyama, R. Ukai, S. C. Armstrong, C. Sornphiphatphong, T. Kaji, S. Suzuki, J. Yoshikawa, H. Yonezawa, N. C. Menicucci, and A. Furusawa,
\emph{Ultra-large-scale continuous-variable cluster states multiplexed in the time domain}, \href{https://doi.org/10.1038/nphoton.2013.287}{Nature Photonics \textbf{7},982-986 (2013)}.

\bibitem{pfistercluster}
M. Chen, N. C. Menicucci, and O. Pfister,
\emph{Experimental Realization of Multipartite Entanglement of 60 Modes of a Quantum Optical Frequency Comb},
\href{https://doi.org/10.1103/PhysRevLett.112.120505}{Phys. Rev. Lett. \textbf{112}, 120505 (2014)}.

\bibitem{trepscluster}
Y. Cai, J. Roslund, G. Ferrini, F. Arzani, X. Xu, C. Fabre, and N. Treps,
\emph{Multimode entanglement in reconfigurable graph states using optical frequency combs},
\href{https://doi.org/10.1038/ncomms15645}{Nature Communications \textbf{8}, 15645 (2017)}.
  
\bibitem{Madsen2022}
Madsen, L.S., Laudenbach, F., Askarani, M.F. et al. 
\emph{Quantum computational advantage with a programmable photonic processor}.  \href{https://doi.org/10.1038/s41586-022-04725-x}{Nature \textbf{606}, 75–81 (2022)}

\bibitem{Boyer2008}
V. Boyer, A. M. Marino, R. C. Pooser, and P. D. Lett, 
\emph{Entangled Images from Four-Wave Mixing}, 
\href{https://doi.org/10.1103/PhysRevA.100.023839}{Science \textbf{321}, 544 (2008)}.

\bibitem{Marino2009}
A. M. Marino, R. C. Pooser, V. Boyer, and P. D. Lett, 
\emph{Tunable delay of Einstein–Podolsky–Rosen entanglement}, 
\href{https://doi.org/10.1103/PhysRevA.100.023839}{Nature  \textbf{457}, 859 (2009)}.

\bibitem{Lopez2020}
J. P. Lopez, A. M. G. de Melo, and J. W. R. Tabosa, 
\emph{Self-amplifying memory based on multiple cascading four-wave mixing via recoil-induced resonance},
\href{https://doi.org/10.1364/ol.394302}{Optics Letters \textbf{45}, 3490 (2020).}
  
\bibitem{sup_mat}
See Supplemental Material at [URL-will-be-inserted-by-publisher], which also includes Refs. \cite{Schilke2011, Bloch_messiah, HomoHetero, BarbosaPRA, Wieman1976, Duan2000}

\bibitem{Schilke2011}
A. Schilke and C. Zimmermann and P. W. Courteille and W. Guerin, \emph{Optical parametric oscillation with distributed feedback in cold atoms}, \href{https://doi.org/10.1038/nphoton.2011.320}{Nature Photonics \textbf{6}, 101-104 (2011)}.

\bibitem{Bloch_messiah}
S. L. Braunstein, 
\emph{Squeezing as an irreducible resource},
\href{https://doi.org/10.1103/PhysRevA.71.055801}{Phys. Rev. A \textbf{71}, 055801 (2005).}

\bibitem{HomoHetero}
H. P. Yuen and V. W. Chan, \emph{Noise in homodyne and heterodyne detection}, \href{https://doi.org/10.1364/OL.8.000177}{Opt. Lett. \textbf{8}, 177-179 (1983).}

\bibitem{BarbosaPRA}
F. A. S. Barbosa, A. S. Coelho, K. N. Cassemiro, P. Nussenzveig, C. Fabre, A. S. Villar, and M. Martinelli, \emph{Quantum state reconstruction of spectral field modes: homodyne and resonator detection schemes}, \href{https://doi.org/10.1103/PhysRevA.88.052113}{Phys. Rev. A \textbf{88}, 052113 (2013).}

\bibitem{Wieman1976}
C. Wieman, T. W. H{\"a}nsch, Physical Review Letters, \emph{Doppler-free laser polarization spectroscopy}, \href{https://doi.org/10.1103/PhysRevLett.36.1170}{Physical Review Letters \textbf{36}, 1170, (1976)}.

\bibitem{Duan2000}
Lu-Ming Duan, G. Giedke, J. I. Cirac, and P. Zoller, \emph{Inseparability Criterion for Continuous Variable Systems}, 
\href{https://link.aps.org/doi/10.1103/PhysRevLett.84.2722}{Phys. Rev. Lett. \textbf{84}, 2722 (2000).}

\bibitem{Allan2016}
A. J. F. de Almeida, M.-A. Maynard, C. Banerjee, D. Felinto, F. Goldfarb, and J. W. R. Tabosa, 
\emph{Nonvolatile optical memory via recoil-induced resonance in a pure two-level system}, 
\href{https://doi.org/10.1103/PhysRevA.94.063834}{Phys. Rev. A \textbf{94}, 063834 (2016).}

\bibitem{Simon2000}
R. Simon, \emph{Peres-horodecki separability criterion for continuous variable systems}, 
\href{https://doi.org/10.1103/PhysRevLett.84.2726}{Phys. Rev. Lett. \textbf{84}, 2726 (2000).}

\bibitem{FBarbosa2011}
F. A. S. Barbosa, A. J. de Faria, A. S. Coelho, K. N. Cassemiro, A. S. Villar, P. Nussenzveig, and M. Martinelli, 
\emph{Disentanglement in bipartite continuous-variable systems}, 
\href{https://doi.org/10.1103/PhysRevA.84.052330}{Phys. Rev. A \textbf{84}, 052330 (2011).}

\bibitem{Xiao2009}
Y. Zhang, U. Khadka, B. Anderson, and M. Xiao,
\emph{Temporal and Spatial Interference between Four-Wave Mixing and Six-Wave Mixing Channels}
\href{https://doi.org/10.1103/PhysRevLett.102.01360}
{Phys. Rev. Lett. \textbf{102}, 013601 (2009).}
 
\end{thebibliography}
\end{document}